\def\Title{Soundness and completeness of quantum root-mean-square errors}
\def\Author{Masanao Ozawa}
\def\Affiliation{Graduate School of Informatics,
Nagoya University, Chikusa-ku, Nagoya, 464-8601, Japan\\  }
\def\Email{ozawa@is.nagoya-u.ac.jp}
\documentclass[twocolumn,showkeys]{revtex4-1}
\usepackage{amsmath,amssymb,amsthm} 
\usepackage{color}
\usepackage{url}
\newcommand{\beq}{\begin{equation}}
\newcommand{\eeq}{\end{equation}}
  \newcommand{\beql}[1]{\begin{equation}\label{eq:#1}}
  \newcommand{\beqa}{\begin{eqnarray}}
  \newcommand{\eeqa}{\end{eqnarray}}
  \newcommand{\beqas}{\begin{eqnarray*}}
  \newcommand{\eeqas}{\end{eqnarray*}}
  \newcommand*{\C}{\mathbb{C}}
 \newcommand*{\R}{\mathbb{R}}
  \newcommand{\M}{{\bf M}}
  \newcommand*{\bE}{\mathbf{E}}
 \newcommand*{\bM}{\mathbf{M}}
  \newcommand*{\bS}{\mathbf{S}}
  \newcommand*{\bx}{\mathbf{x}}
  \newcommand*{\cH}{\mathcal{H}}
  \newcommand*{\cK}{\mathcal{K}}
  \newcommand*{\cM}{\mathcal{M}}

  \newcommand*{\da}{\dagger}
  \newcommand*{\de}{\delta}
  \newcommand{\ep}{\varepsilon}
  \newcommand*{\et}{\eta}
  \newcommand*{\ga}{\gamma}

  \newcommand*{\mb}{\mbox}
  \newcommand*{\nn}{\nonumber}
  \newcommand*{\om}{\omega}
  \newcommand*{\ph}{\phi}
  \newcommand*{\ps}{\psi} 
  
  \newcommand*{\si}{\sigma} 
  \newcommand*{\ta}{\tau}
 \renewcommand*{\th}{\theta}

  \newcommand*{\Eq}[1]{Eq.~(\ref{eq:#1})}

  \newcommand*{\Om}{\Omega}

  \newcommand*{\Ps}{\Psi}                                            
                                            
 \newcommand*{\Then}{\Rightarrow}
  \newcommand*{\Th}{\Theta}                                          
  
  \newcommand*{\eq}[1]{(\ref{eq:#1})}
\newcommand*{\bra}[1]{\langle#1|}
\newcommand*{\ket}[1]{|#1\rangle}

\newcommand*{\bracket}[1]{\langle#1\rangle}
\newcommand*{\ketbra}[1]{\ket{#1}\bra{#1}}
\renewcommand{\And}{\wedge}
  \newtheorem{Theorem}{Theorem}   
  \newenvironment{Proof}{\begin{trivlist}
    \item[\hskip \labelsep {\em \indent Proof.}]}{\qed\end{trivlist}}
\newcommand{\benum}{\begin{enumerate}[{\rm (i)}]\itemsep=0in \parskip=0pt}
\newcommand{\eenum}{\end{enumerate}}
\newcommand{\epg}{\ep_{G}}
\newcommand{\epn}{\ep_{NO}}
\newcommand{\epu}{\overline{\ep}}

\renewcommand{\t}{\tau}
\newcommand{\av}[1]{\langle #1 \rangle}
\renewcommand{\Re}{{\rm Re}}

\newcommand{\bmat}{\left[\begin{array}{rr}}
\newcommand{\emat}{\end{array}\right]}
\newcommand{\bvec}{\left[\begin{array}{r}}
\newcommand{\evec}{\end{array}\right]}
	
\begin{document}
\title{\Large\Title}
\author{\sc\Author}
\email{\Email}
\affiliation{\Affiliation}

\begin{abstract}
{\bf Defining and measuring the error of a measurement is one of the most
fundamental activities in experimental science.
However,  quantum theory shows a peculiar difficulty in extending 
the classical notion of root-mean-square (rms) error to quantum measurements.
A straightforward generalization based on the noise-operator 
was used to reformulate Heisenberg's uncertainty relation
on the accuracy of simultaneous measurements to be universally valid 
and made the conventional formulation testable to observe its violation.   
Recently, its reliability was examined based on an anomaly 
that the error vanishes for some inaccurate measurements, 
in which the meter does not commute with the measured observable.
Here, we propose an improved definition for a quantum generalization 
of the classical rms error, which is state-dependent, operationally definable,
and perfectly characterizes accurate measurements. 
Moreover, it is shown that  the new notion maintains 
the previously obtained universally valid uncertainty relations 
and their experimental confirmations without changing their forms 
and interpretations, in contrast to a prevailing view that a state-dependent 
formulation for measurement uncertainty relation is not tenable. }
\end{abstract}
\pacs{03.65.Ta, 06.20.Dk, 03.67.-a}
\keywords{quantum measurement,
uncertainty principle, 
uncertainty relation,
simultaneous measurement, 
Gauss,
root-mean-square, 
error, 
disturbance,
quantum perfect correlation,
accurate measurement
}
\maketitle

\section*{\bf  Introduction}
The notion of the  mean error of a measurement of  a classical physical quantity 
was first introduced by Laplace \cite[p.~324]{Laplace1812}
as the mean of the absolute value of the error.  Subsequently, the root-mean-square (rms) error was 
introduced by Gauss \cite[p.~39]{Gauss1821} as a mathematically more tractable definition
to derive the principle of the least square,
and has been broadly accepted as the standard definition for the mean error of a measurement.
In those approaches the error of a measurement of a quantity $\Th$ is defined 
as $N=\Om-\Th$,
where $\Om$ is the quantity actually observed, here we call the {\em meter quantity}.  
Then Gauss's rms error is defined as $\av{N^2}^{1/2}$,
where $\av{\cdots}$ stands for the mean value,
while Laplace's mean error as $\av{|N|}$.
From the above definition, Gauss's rms error 
$\epg$ is determined by the joint probability distribution 
\beq
\mu(\th,\om)=\Pr\{\Th=\th,\Om=\om\}
\eeq 
of $\Th$ and $\Om$ as
\beq
\ep_{G}(\mu)^2=\sum_{\om,\th}(\om-\th)^2\, \mu(\th,\om),\label{eq:G}
\eeq
so that $\epg(\mu)=\av{N^2}^{1/2}$,
and it perfectly characterizes accurate measurements:
 $\ep_{G}(\mu)=0$ 
if and only if $\Om=\Th$ holds with probability 1, i.e.,
$\sum \{\mu(\th,\om)\mid \th=\om\}=1$.

A straightforward generalization of Gauss's definition 
to quantum measurements has been introduced as follows 
\cite{Ish91,91QU,BK92}. 
Let $A$ be an observable of a system $\bS$, described by a Hilbert space $\cH$, 
to be measured by a measuring process $\bM$.
Let $M$ be an observable representing the meter of the observer
in the environment $\bE$ described by a Hilbert space $\cK$. 
The Hilbert spaces $\cH$ and $\cK$ are supposed to be finite dimensional
throughout the present paper for simplicity of the presentation, 
although the arguments supporting the main results are extended 
to the infinite dimensional case with well-known mathematical methods.
The time evolution of the total system $\bS+\bE$ during the measuring interaction
with the total Hamiltonian $H$ 
determines the Heisenberg operators $A(0)$,  $M(\t)$ with $0<\t$, where 
 \begin{eqnarray}
 A(0)&=&A\otimes I,\\
 M(\t)&=&U(\t)^{\dagger}(I\otimes M) U(\t),\\  
U(\t)&=&\exp (-i \t H/\hbar).
\end{eqnarray}
To obtain the outcome $\bx$ of this measurement 
the observer measures the observable $M(\t)$
(i.e.,  measures the meter observable $M$ just after the interaction),
instead of measuring $A(0)$
(i.e.,  measuring $A$  just before the the interaction).
The error of this measurement is naturally identified with the observable, 
called the {\em noise operator}, defined by 
\beq
N(A,\bM)=M(\t)-A(0)
\eeq
\cite{AK65,AG88}.
Let $\ket{\psi}$ and $\ket{\xi}$ be the initial states of $\bS$ and $\bE$, 
respectively.
The {\em noise-operator based quantum 
root-mean-square (q-rms) error} of this measurement 
is defined as 
\beql{noise}
\epn(A,\bM,\ket{\psi})=\av{\psi,\xi|N(A,\bM)^2|\psi,\xi}^{1/2},
\eeq
where $\ket{\psi,\xi}=\ket{\psi}\ket{\xi}$ 
 \cite{Ish91,91QU,BK92}. 

This notion was used to reformulate Heisenberg's uncertainty relation
for the accuracy of simultaneous measurements
to be universally valid 
\cite{03UVR,03HUR,03UPQ,04URN,04URJ,Hal04,WHPWP13,Bra13,Bra14,14EDR} 
and made the conventional formulation testable to observe its violation.
\cite{12EDU,RDMHSS12,13EVR,RBBFBW14,13VHE,14A1,16A3}.   

Recently, Busch, Lahti, and Werner (BLW) \cite{BLW14RMP} raised 
a reliability problem for quantum generalizations of the classical rms error,
comparing the noise-operator based q-rms error 
with the Wasserstein 2-distance, another error measure 
based on the distance between probability measures,
and pointed out several discrepancies between those two error measures
in favor of the latter.

In order to resolve the conflict, here we introduce the following requirements for 
any  sensible error measure generalizing the classical root-mean-square error:
(I) the operational definability, 
(II) the correspondence principle, 
(III) the soundness, and 
(IV) the completeness.
The operational definability ensures that the error measure is definable by the operational
description of the measuring process. 
The correspondence principle ensures that the error measure is consistent 
with the classical rms error in the case when the latter is also applicable.
The soundness ensures that the error measure vanishes for
any accurate measurements, while the completeness ensures that 
the error measure does not vanish for any inaccurate measurements.
As shown later,  the noise-operator based q-rms error $\epn$ satisfies 
all the requirements (I)--(III) except (IV), 
whereas any error measures based on the distance 
of probability measures, such as the Wasserstein 2-distance, 
satisfy (I) and (III) but do not satisfy (II) nor (IV).
We propose an improved definition for a quantum generalization 
of the classical rms error, which is still based on the noise operator but
satisfies all requirements (I)--(IV).
Moreover, it is shown that  the new error measure maintains 
the previously obtained universally valid uncertainty relations 
\cite{03UVR,03HUR,03UPQ,04URN,04URJ,Hal04,WHPWP13,Bra13,Bra14,14EDR} 
and their experimental confirmations 
\cite{12EDU,RDMHSS12,13EVR,RBBFBW14,13VHE,14A1,16A3}
without changing their forms 
and interpretations, in contrast to a prevailing view that a state-dependent 
formulation for measurement uncertainty relation is not tenable
\cite{BLW14RMP,DN14,KJR14}. 

\section*{\bf  Results}
\paragraph*{\bf Operational definability.}  
The probability distribution of the output $\bx$ of the measurement 
is given by
\beq
\Pr\{\bx=x\|\ket{\psi}\}
=\bracket{\psi,\xi|P^{M(\t)}(x)|\psi,\xi},
\eeq
where $P^{M(\t)}(x)$ is the spectral projection of $M(\t)$ for $x\in\R$,
i.e., $P^{M(\t)}(x)$ is the projection with range 
$\{\Psi\in \cH\otimes\cK\mid M(\t)\Psi=x\Psi\}$.
It is fairly well-known that every measuring process has its 
probability operator-valued measure (POVM)
that operationally describes the statistics of
the measurement outcome \cite{Hel76,Hol82,Dav76,84QC}.
The {\em POVM} $\Pi$ of the measuring process $\bM$ is 
a family  $\Pi=\{\Pi(x)\}_{x\in\R}$ 
of positive operators on $\cH$ defined by 
\beq
\Pi(x)=\langle\xi| P^{M(\t)}(x)|\xi\rangle,
\eeq
and satisfies the generalized Born formula
\beq
\Pr\{\bx=x\|\ket{\psi}\}=\bracket{\psi|\Pi(x)|\psi}.
\eeq

We consider the requirements for any quantum 
generalization  $\ep$ of the classical root-mean-square error $\epg$
to quantify the mean error $\ep(A,\bM,\ket{\psi})$ 
of the measurement of an observable $A$ 
in a state $\ket{\psi}$ described by a measuring process ${\bf M}$;
we shall also write $\ep(A,\bM,\rho)$ if the state is represented by
a density operator $\rho$.
The first requirement is formulated using the notion of POVM
as follows.
\medskip

 (I)  {\em Operational definability.} {The error measure  $\ep$ should be 
definable by the POVM $\Pi$ of the measuring process $\bM$, 
the observable $A$ to be measured, and the initial state $\ket{\psi}$ of the 
measured system $\bS$.}
\medskip

The operational definability determines the mathematical domain of
the error measure and requires that the mean error (i.e, the value of the
error measure in the given state) should be determined by the operational
description of the statistics of measurement outcomes.

The {\em n-th moment operator} $\hat{\Pi}^{(n)}$ of the POVM $\Pi$ is
defined by
\beq
\hat{\Pi}^{(n)}=\sum_{x}x^{n} \Pi(x).
\eeq
We write $\hat{\Pi}=\hat{\Pi}^{(1)}$.
Then the relation
\beqa
\epn(A,\bM,\ket{\psi})^{2}
&=&\Re \av{\psi|A^2-2A\, \hat{\Pi}+\hat{\Pi}^{(2)}|\psi}\nn\\
\label{eq:RMSE-POVM}
\eeqa
holds \cite[Theorem 4.5]{04URN}.

Thus, $\epn$ can be defined by the observable $A$, the POVM $\Pi$, 
and the state $\ket{\psi}$, so that it satisfies the operational definability.
In what follows, we shall write  $\epn(A,\Pi,\ket{\psi})=\epn(A,\bM,\ket{\psi})$ 
if $\Pi$ is the POVM of $\bM$.

\medskip \paragraph*{\bf Correspondence principle.}
The second requirement is based on a common practice in generalizing a
classical notion to quantum mechanics.  
Even in quantum mechanics, there are cases where the original classical notions
are directly applicable, and in those cases the generalized notions should 
be consistent with the original ones.

In the problem of generalizing the classical root-mean-square error
to quantum mechanics, this principle is applied to the case where
$A(0)$ and $M(\t)$ commute as two operators.  
In this case, the observables $A(0)$ and $M(\t)$ are
jointly measurable and their joint probability distribution
$\mu(x,y)$ is given by
\beq
\mu(x,y)=\av{\psi,\xi|P^{A(0)}(x)P^{M(\t)}(y)|\psi,\xi}.
\eeq
Then we can apply the classical definition of the root-mean-square error 
to the joint probability distribution $\mu$ to obtain the classical 
root-mean-square error $\ep_G(\mu)$ of this measurement;
in this case, the measuring process is classically described as a black-box 
with the input-output joint probability distribution $\mu(x,y)$. 
Thus, the quantum generalization $\ep$ should satisfy
\beql{CP}
\ep(A,\bM,\ket{\psi})=\ep_G(\mu).
\eeq

Thus, we should require that \Eq{CP} holds if $A(0)$ and $M(\t)$ commute.  
However,  we should proceed further to avoid possible inconsistencies,
since there is a case where a pair of observables commute only on a subspace
and they have the joint probability distribution only for states in that
subspace as discussed by von Neumann \cite[p.~230]{vN55}.
To include such a general situation, we define the notions of 
commutativity and joint probability
distribution in a sate-dependent manner.
We say that observables $X$ and $Y$ {\em commute} in a state $\ket{\Psi}$ if  
\beql{com}
P^{X}(x)P^{Y}(y)\ket{\Psi}=P^{Y}(y)P^{X}(x)\ket{\Psi}
\eeq
for any $x,y$.
A probability distribution $\mu(x,y)$ on $\R^{2}$, i.e., $ \mu(x,y)\ge 0$ and
$\sum_{x,y}  \mu(x,y)=1$,
is called a {\em joint probability distribution (JPD)} of observables
$X,Y$ in $\ket{\Psi}$ if
\beql{JPD}
\av{\Ps|f(X,Y)|\Ps}=\sum_{x,y}f(x,y)\,\mu(x,y)
\eeq
for any polynomial $f(X,Y)$ of observables $X,Y$.
Then, there exists a JPD of observables $X,Y$ in $\ket{\psi}$ 
if and only if $X$ and $Y$ commute in $\ket{\psi}$ as shown in 
Theorem 1 in Methods:
State-dependent commutativity and joint probability distributions. 
In this case, the JPD $\mu$ is uniquely determined by 
\beq
\mu(x,y)=\av{\Ps|P^{X}(x)P^{Y}(y)|\Ps}.
\eeq

To prevent the inconsistency between the original classical notion and
its quantum generalization we pose the following requirement.   
\medskip

(II)  {\em Correspondence principle.}
{In the case where $A(0)$ and $M(\t)$ commute 
in the initial state $\ket{\psi,\xi}$, then the relation
\beq
\ep(A,\bM,\ket{\psi})=\epg(\mu)
\eeq
holds for the JPD $\mu$ of  $A(0)$ and $M(\t)$
in $\ket{\psi,\xi}$.}
\medskip

Suppose that $A(0)$ and $M(\t)$ commute in  $\ket{\ps,\xi}$.
Let $\mu$ be their JPD  in  $\ket{\ps,\xi}$.
From  Eqs.~\eq{G}, \eq{noise}, and \eq{JPD},  we have
\beq
\epn(A,\bM,\ket{\psi})=\epg(\mu).
\eeq
Thus, the noise-operator based q-rms error $\epn$ 
satisfies the correspondence principle.
\medskip 

\paragraph*{\bf Soundness.}
To discuss the soundness we need to clarify what measuring process 
${\bf M}$ is considered to accurately measure an observable $A$ 
in a given state $\ket{\psi}$.
This fundamental problem has, to the best of our knowledge, 
not been discussed in the literature
except for our previous investigations 
\cite{05PCN,06QPC,11QRM,16A2}, 
in which we introduced the following definition.
We say that the measuring process $\bM$ {\em accurately measures} 
an observable $A$ in a state $\ket{\psi}$ if $A(0)$ and $M(\t)$ 
commute in $\ket{\psi,\xi}$ and their JPD $\mu$ satisfies 
$\mu(A(0)=M(\t))=1$, where $\mu(A(0)=M(\t))=\sum_{x,y:x=y}\mu(x,y)$.
We will provide a justification of this definition including 
its operational accessibility in Methods: State-dependent 
definition for accurate measurements of quantum observables.

Under the above definition we pose the soundness requirement.
\medskip

(III)  {\em Soundness.}  {The error measure $\ep$ vanishes 
for any accurate measurements.}
\medskip

Now, we can see that any error measure $\ep$ satisfying 
the correspondence principle, (II), also satisfies the soundness, (III),
since in this case we have the JPD $\mu$ of $A(0)$ and $M(\t)$ 
satisfying  $\ep(A,\M,\ket{\psi})=\epg(\mu)=0$.

Since the noise-operator based rms error $\epn$ satisfies the correspondence
principle, (II), it also satisfies the Soundness, (III).

Note that if $A$ is accurately measured in $\ket{\psi}$,
then $A$ and $\Pi$ are identically distributed in $\ket{\psi}$.

\medskip

\paragraph*{\bf Completeness.}
Now, we introduce the following requirement.
\medskip

 (IV)  {\em Completeness.} The measurement is accurate if the error measure $\ep$
vanishes.
\medskip

Busch,  Heinonen, and Lahti \cite[p.~263]{BHL04} pointed out that there is
a measuring process $\M$ such that $\epn(A,\M,\ket{\psi})=0$ 
but $\bM$ does not accurately measure $A$ in $\ket{\psi}$. 
For a simple example,  let 
\beql{ex}
A=
\bmat
1&1\\
1&1
\emat,\
M=\bmat
1&1\\
1&-1
\emat,\
\ket{\psi}=\bvec
1\\
0
\evec
\eeq
with $\Pi(y)=P^{M}(y)$.
Then we have $\ep_{NO}(A,\Pi,\ket{\psi})=0$,
but the measurement is not accurate, 
since $A$ and $\Pi$ are not identically distributed as
$\av{\psi|P^{A}(2)|\psi}=1/2$ 
but $\av{\psi|\Pi(2)|\psi}=0$.

Thus, the noise-operator based q-rms error $\ep_{NO}$ does not satisfy the 
completeness requirement.
As shown above,  the noise-operator based q-rms error $\epn$ satisfies 
all the requirements (I)--(III) but does not satisfy (IV). 
\medskip 

\paragraph*{\bf Locally uniform quantum root-mean-square error.}
We call any error measure $\ep$ satisfying (I) and (II) 
a {\em quantum root-mean-square (q-rms) error}.
A q-rms error $\ep$ is said to be {\em sound} if it satisfies (III).
It is said to be {\em complete} if it satisfies (IV).
A sound and complete error measure correctly 
indicates the cases where the measurement is accurate and where not
\cite[p.~1263]{BLW14RMP}.
A primary purpose of this paper is to find a sound and complete q-rms error, 
and to establish universally valid uncertainty relations based on it.

We shall show that there is a simple method to strengthen the
noise-operator based q-rms error to obtain a sound and complete q-rms error.
In addition to (I)--(IV), this error measure is shown 
to have the following two properties.
\medskip

(V)  {\em Dominating property.} The error measure $\ep$ 
dominates the noise-operator based q-rms error $\epn$, 
i.e., $\epn(A,\Pi,\ket{\psi})\le{\ep}(A,\Pi,\ket{\psi})$ for all $A,\Pi,\ket{\psi}$.
\medskip

(VI) {\em Conservation property for dichotomic measurements.}
The error measure $\ep$ coincides with the noise-operator based q-rms error 
$\epn$ for dichotomic 
measurements, i.e., $\epn(A,\Pi,\ket{\psi})={\ep}(A,\Pi,\ket{\psi})$ 
if $A^2=\hat{\Pi}^{(2)}=I$.
\medskip

For any ${t\in\R}$, define 
\beq
{\ep_t(A,\Pi,\ket{\psi})=\ep_{NO}(A,\Pi,e^{-itA}\ket{\psi} ).}
\eeq
We call ${\{\ep_t(A,\Pi,\ket{\psi})\}_{t\in\R}}$ the 
{\em q-rms error  profile} 
for $A$ and $\Pi$ in $\ket{\psi}$.
If $A(0)$ and $M(\t)$ commute in the state $\ket{\psi,\xi}$,
then we have 
\beql{ept-epn}
\ep_t(A,\Pi,\ket{\psi})=\epn(A,\Pi,\ket{\psi})
\eeq
 for all $t\in\R$.
Thus, the q-rms error  profile is considered to provide additional information
about the error of measurement $\bM$ in the case where $A(0)$ and $M(\t)$ 
do not commute in the state $\ket{\psi,\xi}$.

To obtain a numerical error measure from ${\{\ep_t(A,\Pi,\ket{\psi})\}_{t\in\R}}$, 
we define the {\em  locally uniform q-rms error} by
\beq
\epu(A,\Pi,\ket{\psi})=\sup_{t\in\R}\ep_t(A,\Pi,\ket{\psi}).
\eeq

Then $\epu$ is a sound and complete q-rms error, satisfying both
the dominating property, (V), and the conservation property for dichotomic
measurements, (VI), as shown in Theorem 3 in Methods:
Sound and complete quantum root-mean-square errors,
where we introduce other two sorts of q-rms errors to clarify the
physical motivation behind the above definition.

For the example given in \Eq{ex}, we have 
\beq
\ep_t(A,\Pi,\ket{\psi})=2|\sin t|, \quad\mb{and}\quad  \epu(A,\Pi,\ket{\psi})=2,
\eeq
despite of  the relation $\ep_{NO}(A,\Pi,\ket{\psi})=0$,
the relation $\epu(A,\Pi,\ket{\psi})=2$ correctly indicate that
the measurement of $A$ described by \Eq{ex} is not 
an accurate measurement.
\medskip

\section*{\bf  Discussion}

\paragraph*{\bf Wasserstein 2-distance.}
 In what follows, we shall show that 
the Wasserstein 2-distance
satisfies the operational definability, (I), and the soundness, (III),
but does not satisfy the correspondence principle, (II), nor 
the completeness, (IV).

Let $\mu^{A}_{\ket{\psi}}$ and $\mu^{\Pi}_{\ket{\psi}}$ 
be the probability distributions of 
$A$ and $\Pi$ in state $\ket{\psi}$, i.e., 
\beqa
\mu^{A}_{\ket{\psi}}(x)&=&\av{\psi|P^{A}(x)|\psi},\\
\mu^{\Pi}_{\ket{\psi}}(y)&=&\av{\psi|\Pi(y)|\psi}.
\eeqa
BLW \cite{BLW14RMP} advocated 
the Wasserstein 2-distance $W_2(\mu^{A}_{\ket{\psi}},\mu^{\Pi}_{\ket{\psi}})$
between $\mu^{A}_{\ket{\psi}}$ and $\mu^{\Pi}_{\ket{\psi}}$ as an alternative 
quantum generalization of the classical rms error
in comparison with the noise-operator based q-rms error
$\epn(A,\Pi,\ket{\psi})$.  
The {\em Wasserstein 2-distance} is defined as 
\beq
W_2(\mu^{A}_{\ket{\psi}},\mu^{\Pi}_{\ket{\psi}})=
\inf_{\ga} \epg(\ga),
\eeq
where the infimum is taken over all the probability distributions
$\ga(x,y)$ on $\R^2$
such that $\ga(x,\R)=\mu^{A}_{\ket{\psi}}(x)$ and 
$\ga(\R,y)=\mu^{\Pi}_{\ket{\psi}}(y)$.
Thus, $W_2(\mu^{A}_{\ket{\psi}},\mu^{\Pi}_{\ket{\psi}})$ satisfy  the operational
definability, (I).
It should be pointed out that the Wasserstein 2-distance 
$W_2(\mu^{A}_{\ket{\psi}},\mu^{\Pi}_{\ket{\psi}})$ does not satisfy 
the correspondence principle, (II).  
To see this,
suppose that $A(0)$ and $M(\t)$ commute in $\ket{\psi,\xi}$.
In this case, we have
\beq
\ep_{G}(\mu)^2=\si(A(0))^2+\si(M(\t))^2-
2{\rm Cov}+({\rm Bias})^2,\\
\eeq
where ${\rm Cov}=\bracket{(A(0)-a)(M(\ta)-m)}$,
${\rm Bias}=a-m$,  $a=\av{A(0)}$, and $m=\av{M(\ta)}$.
The JPD $\mu(x,y)$ always satisfies the condition that  
$\mu(x,\R)=\mu^{A}_{\ket{\psi}}(x)$ and 
$\mu(\R,y)=\mu^{\Pi}_{\ket{\psi}}(y)$.
Thus, we have
\beql{W-G}
W_2(\mu^{A}_{\ket{\psi}},\mu^{\Pi}_{\ket{\psi}})
\le \ep_{G}(\mu).
\eeq

For the case where $\mu^{A}_{\ket{\psi}}=\mu^{\Pi}_{\ket{\psi}}$,
we have $W_2(\mu^{A}_{\ket{\psi}},\mu^{\Pi}_{\ket{\psi}})=0$, but
$\ep_{G}(\mu)=0$ only if $\mu(A(0)=M(\t))=1$.
To consider a typical case where $\ep_{G}(\mu)>0$, suppose that 
$A(0)$ and $M(\t)$ are independent.  Then we have
$\si(A(0))=\si(M(\t))$, ${\rm Cov}=0$, and ${\rm Bias}=0$,
and hence $\ep_{G}(\mu)=\sqrt{2}\si(A)$.
Thus,  $\ep_{G}(\mu)>0$ whenever $\si(A)>0$.
For instance, let 
\beqa
A&\!=\!&\ketbra{0}+\ketbra{1}-\ketbra{2}-\ketbra{3},\\
M&\!=\!&\ketbra{0}-\ketbra{1}+\ketbra{2}-\ketbra{3},\\
A(0)&\!=\!&A\otimes I,\\
M(\t)&\!=\!&U^{\da}(I\otimes M)U=M\otimes I,\\
\ket{\psi}&\!=\!&(\ket{0}+\sqrt{2}\ket{1}+\sqrt{2}\ket{2}+2\ket{3})/3.
\eeqa
Then, we have the joint probability distribution $\mu$ for $A(0)$ and $M(\t)$
in $\ket{\psi,\xi}$ for arbitrary $\ket{\xi}$ such that 
\beqa
\mu(+1,+1)=1/9, &\quad &
\mu(+1,-1)=2/9,\\
\mu(-1,+1)=2/9,& &
\mu(-1,-1)=4/9.
\eeqa
We have
$\mu^{A}_{\ket{\psi}}(+1)=\mu^{\Pi}_{\ket{\psi}}(+1)=1/3$, 
$\mu^{A}_{\ket{\psi}}(-1)=\mu^{\Pi}_{\ket{\psi}}(-1)=2/3$, and 
$A(0)$ and $M(\t)$ are independent.
Thus we have $W_2(\mu^{A}_{\ket{\psi}},\mu^{\Pi}_{\ket{\psi}})=0$, but
$\si(A)=2\sqrt{2}/3$ and $\epg(\mu)=\epn(A,\Pi,\ket{\psi})=4/3$. 
Thus, the Wasserstein 2-distance does not satisfy the correspondence principle, (II).

Note that the above example also shows that the Wasserstein 2-distance 
$W_2(\mu^{A}_{\ket{\psi}},\mu^{\Pi}_{\ket{\psi}})$
does not satisfy the completeness, (IV),
whereas it satisfies the soundness, (III), since 
$\epg(\mu)=0$ holds in \Eq{W-G}
for any accurate measurement.

The logical relationships among requirements (I)--(IV) are summarized as 
follows. Under the the major premise (I), we have shown that (i) (III) follows from (II), 
(ii) (II) does not follow from (III), since the Wasserstein 2-distance satisfies (III) but does not satisfies (II),
and that (iii) (II) and (IV) are independent, since $\ep_{NO}$ satisfies (II) but does not satisfy (IV)
and since there exists an error measure $\ep$ satisfying (IV) but not satisfying (II), e.g. $\ep(A,\Pi,\ket{\psi})=\sup_{\ket{\phi}}\ep_{NO}(A,\Pi,\ket{\phi})$, where $\ket{\phi}$ 
varies over all the states.
Note that there exists an error measure $\ep$ satisfying (I), (III), and (IV), but does not
satisfy (II), e.g., $\ep(A,\Pi,\ket{\psi})=\sup_{\ket{\phi}}\ep_{NO}(A,\Pi,\ket{\phi})$,
where $\ket{\phi}$ varies over the cyclic subspace $\mathcal{C}(A,\ket{\psi})$ generated
by $A$ and $\ket{\psi}$ \cite{ 06NDQ-}. 
\medskip 

\paragraph*{\bf Universally valid uncertainty relations.}
In what follows, we shall show that 
all the universally valid measurement
uncertainty relations obtained so far 
\cite{03UVR,03UPQ,04URN,Hal04,WHPWP13,Bra13,Bra14}
for the noise-operator based q-rms error
are maintained with the same forms by property (V)
and that their experimental confirmations reported so far 
\cite{12EDU,RDMHSS12,13EVR,RBBFBW14,13VHE,14A1,16A3}
for dichotomic measurements
are also reinterpreted to confirm the relations for the new error measure
by property (VI).
Moreover, the state-independent 
formulation based on this notion maintains
Heisenberg's original form for the measurement uncertainty relation,
whereas the state-dependent formulation violates it.
The new error measure thus clears a prevailing view
that the state-dependent formulation of measurement 
uncertainty relations is not tenable
\cite{BLW14RMP,DN14,KJR14}.

Let $A,B$ be two observables of a quantum system $\bS$
described by a Hilbert space $\cH$.
Any simultaneous measurement of $A$ and $B$
in a state $\ket{\psi}$ defines a joint POVM $\Pi(x,y)$ on $\R^2$
for the Hilbert space $\cH$, for which the marginal POVM
$\Pi_A(x)=\Pi(x,\R)$ describes the $A$-measurement and 
the marginal POVM $\Pi_B(y)=\Pi(\R,y)$ describes 
the $B$-measurement \cite{04URJ}.
Then the mean  errors of the simultaneous measurement
of $A$ and $B$ described by the joint POVM $\Pi(x,y)$ in the state
$\ket{\psi}$ are defined as $\ep(A,\Pi_A,\ket{\psi})$ and $\ep(B,\Pi_B,\ket{\psi})$,
respectively, for a given q-rms error $\ep$.
In what follows we abbreviate $\ep(A)$ to $\ep(A,\Pi_A,\ket{\psi})$ and 
$\ep(B)$ to $\ep(B,\Pi_B,\ket{\psi})$ unless confusion may occur.

\sloppy
The above general formulation includes the error-disturbance relation
for the $A$-measurement error of a measuring process $\bM$
and the thereby caused disturbance on $B$,
since the $B$-disturbance is generally defined by the error of the
accurate $B$-measurement following the $A$-measurement
\cite{03UVR,04URN,BHL07}.
This definition of the $B$-disturbance is described in the Heisenberg
picture as follows.
Given a measuring process $\bM$, we can make
an accurate simultaneous measurement of commuting observables
$M(\t)$ and $B(\t)$.
Then an approximate simultaneous measurement of $A(0)$ and $B(0)$ is obtained
if the measurement of $A(0)$ is replaced by the accurate measurement of $M(\t)$
and the measurement of $B(0)$ is replaced by the  accurate measurement of $B(\t)$.
This simultaneous measurement is described by the joint POVM $\Pi$ defined by
\beq
\Pi(x,y)=\av{\xi|P^{M(\t)}(x)P^{B(\t)}(y)|\xi}.
\eeq
In this case, for a given q-rms error measure $\ep$, we define 
the {\em mean error} $\ep(A,\bM,\ket{\psi})$ of the $A$ measurement carried out by
$\bM$ in $\ket{\psi}$ as $\ep(A,\bM,\ket{\psi})=\ep(A,\Pi_A,\ket{\psi})$
and the the {\em mean disturbance} $\et(B,\bM,\ket{\psi})$ of $B$ caused 
by $\bM$ in $\ket{\psi}$ as $\et(B,\bM,\ket{\psi})=\ep(B,\Pi_B,\ket{\psi})$.
In what follows we abbreviate $\ep(A)$ to $\ep(A,\bM,\ket{\psi})$ and 
$\et(B)$ to $\et(B,\bM,\ket{\psi})$ unless confusion may occur.

As above, any general relation for $\ep(A)$ and $\ep(B)$ implies
a general relation for $\ep(A)$ and $\et(B)$, while any counter example for
a general relation for $\ep(A)$ and $\et(B)$ is also a counter example for 
the corresponding relation for $\ep(A)$ and $\ep(B)$.

In this respect, it should be noted that the recent claim by Korzekwa, Jennings, 
and Rudolph (KJR) \cite{KJR14} of the impossibility of state-dependent
error-disturbance relations is unfounded.  In fact, KJR admitted that their
basic assumption called the operational requirement (RO) can be 
applied to the notion of disturbance, but cannot be applied to the notion 
of error \cite[p.~052108-6]{KJR14}; it can be easily seen that if (RO) were
to be applied to the error, it would contradict the correspondence principle.   
However, such a discrimination between the disturbance and the error
contradicts the above standard definition of the disturbance as the error 
of a successive measurement.

Heisenberg's original formulation of the uncertainty principle
states that canonically conjugate observables $Q,P$  can be measured
simultaneously only with a characteristic constraint \cite[p.~172]{Hei27}
\beql{HUP}
\ep(Q)\ep(P)\ge\frac{\hbar}{2},
\eeq
where the unambiguous lower bound ${\hbar}/{2}$ is 
due to a subsequent elaboration by 
Kennard \cite{Ken27} (see also \cite{15A2}).
Heisenberg justified this relation under the repeatability hypothesis or
its approximate version, an obsolete assumption on the state change 
in measurement; see Ref.~\cite{15A2} for a detailed discussion.

A counter example of Heisenberg's relation \eq{HUP} 
was shown in Ref.~\cite{02KB5E} in the error-disturbance scenario with $\ep=\epn$,
using a position measuring model originally constructed in Ref.~\cite{88MS} 
to invalidate the standard quantum limit for gravitational-wave detectors
with free-mass probe \cite{BVT80,CTDSZ80}.
In Ref.~\cite{13DHE} continuously many linear position measuring 
processes including the above have been constructed that violate Heisenberg's  
relation \eq{HUP} in the error-disturbance scenario for an arbitrary choice of 
the q-rms error $\ep$.
Thus, the violation of Heisenberg's relation \eq{HUP} is not due to a particular
choice of the q-rms error $\ep$.

In contrast to the violation of \Eq{HUP} in the state-dependent formulation, 
 Appleby \cite{App98c} showed the relation 
\beql{SIHUP}
\sup_{\ket{\psi}}\ep(Q,\Pi_Q,\ket{\psi})\,\sup_{\ket{\psi}}\ep(P,\Pi_P,\ket{\psi})\ge\frac{\hbar}{2},
\eeq 
holds for $\ep=\epn$,
except for the case where $\sup_{\ket{\psi}}\ep(Q,\Pi_Q,\ket{\psi})= 0$ or 
$\sup_{\ket{\psi}}\ep(P,\Pi_P,\ket{\psi})=0$,
where the supremum is taken over all the possible states $\ket{\psi}$.
An apparent drawback of the above relation is that 
the state-independent error measure $\sup_{\ket{\psi}}\ep(Q,\Pi_Q,\ket{\psi})$ and
$\sup_{\ket{\psi}}\ep(P,\Pi_P,\ket{\psi})$ are defined by a q-rms error $\ep$ that is not
complete. 
However, this drawback turns out to be immediately cleared if one uses  
the uniform q-rms error $\epu$ instead, since the relation 
\beq
\sup_{\ket{\psi}}\epn(X,\Pi_X,\ket{\psi})=\sup_{\ket{\psi}}\epu(X,\Pi_X,\ket{\psi})
\eeq 
holds obviously for any observable $X$.
Thus,  \Eq{SIHUP} holds for $\ep=\epu$, one of the sound and complete q-rms errors.
It should be noted that in the state-independent formulation as above 
the error measures
$\sup_{\ket{\psi}}\ep(Q,\Pi_Q,\ket{\psi})$ and $\sup_{\ket{\psi}}\ep(P,\Pi_P,\ket{\psi})$
often diverges \cite{App98c,13DHE}.  
Even in the original $\gamma$-ray thought experiment, the error measure
$\sup_{\ket{\psi}}\ep(Q,\Pi_Q,\ket{\psi})$ diverges as the wave packet goes beyond
the scope of the microscope. 
Thus, Heisenberg's original form holds in the state-independent formulation
but not due to the tradeoff between the resolution power and the Compton recoil.
The notion of the resolution power of a microscope is well-defined only in the case
where the object is well-localized in the scope of the microscope, and it cannot be
captured by the state-independent formulation.
The above remarks are also applied to the recent revival of the state-independent
formulation by Busch, Lahti, and Werner \cite{BLW13,BLW14JMP}; in fact,
the Busch-Lahti-Werner formulation in \cite{BLW13} is equivalent 
to Appleby's formulation \cite{App98c} for any linear measurements \cite{13DHE}.
For detailed discussions, we refer the reader to Ref. \cite{13DHE}.

A generalization of Heisenberg's relation \eq{HUP} to arbitrary pair of observables
$A$ and $B$ is obtained by using the noise-operator based rms error 
$\ep=\epn$ as the relation 
\beql{HUPAB}
\ep(A)\ep(B)
\ge C_{A,B},
\eeq
where $C_{A,B}=\frac{1}{2}|\av{\psi|[A,B]|\psi}|$,
holding for any joint POVM's with unbiased or independent 
noise operators \cite{AK65,YH86,AG88,Ish91,91QU,Ray94,03UPQ}
(see also \cite{App98a,App98b}).
By the dominating property,  (V), the above relation also holds for
the uniform rms error $\ep=\epu$. 

Using the noise-operator based q-rms error $\ep=\epn$,
the first universally valid relation
\beqa
\ep(A)\ep(B)+\ep(A)\si(B)+\si(A)\ep(B) 
&\!\!\ge\!\!&   C_{A,B},
\label{eq:UEDR}
\eeqa
was given in 2003 \cite{03UVR,03UPQ}, 
which are universally valid for any observables $A,B$, any system state
$\ket{\psi}$, and any joint POVM $\Pi$,
where the standard deviations $\si(A),\si(B)$ are taken in the state $\ket{\psi}$.
By the dominating property, (V),
the above relation also holds for the uniform q-rms error $\ep=\epu$.
Thus, we have a state-dependent
universally valid uncertainty relation for 
simultaneous measurements described by a sound and 
complete q-rms error.

Using the noise-operator based q-rms error $\ep=\epn$,
Branciard \cite{Bra13,Bra14} considerably strengthened the above universally valid
relation \eq{UEDR} as well as the relations proposed by Hall \cite{Hal04}
and by Weston, Hall, Palsson,  and Wiseman \cite{WHPWP13}
in several ways.
All those Branciard relations also hold for the uniform q-rms error $\ep=\epu$
by the dominating property, (V); 
see Branciard \cite[Section IV]{Bra14} for the alternative forms 
of the above mentioned relations to which the dominating property 
can directly apply.

Those universally valid relations for the noise-operator based q-rms error
have already been experimentally confirmed
in the error-disturbance scenario for dichotomic 
measurements 
(i.e., $A(0)^2=B(0)^2=M(\t)^2=B(\t)^2=I$)
with observing the violation of \Eq{HUPAB}
\cite{12EDU,RDMHSS12,13EVR,RBBFBW14,13VHE,14A1,16A3}.
Interestingly, the above experiments were intended to confirm 
relations for the noise-operator based q-rms error $\ep=\epn$,
but they also can be reinterpreted as confirmations for the corresponding
relations and the violation of \Eq{HUPAB}
with the uniform q-rms error $\ep=\epu$, 
one of sound and complete q-rms errors, 
since in those experiments we have $\epn=\epu$ by the
conservation property for dichotomic measurements, (VI).
Thus, we already have a well-developed theory of state-dependent
measurement uncertainty relations based on 
a sound and complete q-rms error, in contrast to  
a prevailing claim that the state-dependent formulation of
measurement uncertainty relations is not tenable
\cite{BLW14RMP,DN14,KJR14}.

\section*{Methods}
\paragraph*{\bf State-dependent commutativity and joint probability distributions.}
The state-dependent notion of commutativity was originally
discussed by von Neumann \cite[p.~230]{vN55} as follows.
Suppose that $\ket{\Psi}$ is a superposition of common eigenstates of $X$ and $Y$, 
namely, there exists an orthonormal family $\{\ket{x,y}\}$ of states such that 
$X\ket{x,y}=x\ket{x,y}$,  
$Y\ket{x,y}=y\ket{x,y}$, 
and that $\ket{\Psi}=\sum_{x,y}\ket{x,y}\av{x,y|\Psi}$.
In this case, a measurement of the observable 
\beq
Z=\sum_{x,y}z_{x,y}\ketbra{x,y}
\eeq
with a one-to-one assignment of real values $(x,y)\mapsto z_{x,y}$ gives a joint measurement
of $X$ and $Y$ in the state $\ket{\Psi}$ and their joint probability distribution 
$\mu(x,y)=\Pr\{X=x,Y=y\}$ of $X$ and $Y$ is given by 
\beq
\mu(x,y)=|\av{\Ps|x,y}|^2=\av{\Ps|P^{X}(x)P^{Y}(y)|\Ps}.
\eeq
In this case, $X$ and $Y$ commute on the subspace $\cM$ spanned by
$\{\ket{x,y}\}$ but do not necessarily commute on $\cM^\perp$. 

Then we have the following theorem.

\begin{Theorem}
For any pair of observables $X,Y$ and state $\ket{\Psi}$, 
the following conditions are all equivalent.

(i) The state $\ket{\Psi}$ is a superposition of common eigenstates of $X$ and $Y$.

(ii) The observables $X$ and $Y$ commute in the state $\ket{\Psi}$, 
i.e., \Eq{com} holds for any $x,y$.

(iii) There exists a JPD $\mu$ of $X$ and $Y$ in $\ket{\Psi}$, i.e., 
there exists a probability distribution $\mu(x,y)$ on $\R^{2}$ satisfying
\Eq{JPD} for any polynomial $f(X,Y)$ of observables $X,Y$.

(iv) $\sum_{x,y}\av{\Psi|P^{X}(x)\And P^{Y}(y)|\Psi}=1$.

 In this case, the JPD $\mu$ is uniquely determined by 
\beql{JPD2}
\mu(x,y)=\av{\Ps|P^{X}(x)P^{Y}(y)|\Ps}.
\eeq
\end{Theorem}
\begin{Proof}
The following proof is obtained by adapting the more general arguments 
previously given in \cite{Gud68,Yli85,06QPC,11QRM,16A2} to the case discussed here.

(i) $\Then$ (iv): Suppose that $\ket{\Psi}$ is a superposition of common eigenstates 
of $X$ and $Y$, namely, there exists an orthonormal family of states $\{\ket{x,y}\}$ 
such that $X\ket{x,y}=x\ket{x,y}$,  $Y\ket{x,y}=y\ket{x,y}$, 
and that $\ket{\Psi}=\sum_{x,y}\ket{x,y}\av{x,y|\Ps}$.
Then we have 
$$
\sum_{x,y}P^{X}(x)\And P^{Y}(y)\ket{\Psi}=\sum_{x,y}\ket{x,y}\av{x,y|\Ps}=\ket{\Psi},
$$
and hence (iv) holds.

(iv) $\Then$ (ii): 
It is easy to see that 
\beqas
P^{X}(u)[P^{X}(x)\And P^{Y}(y)]&=&\de_{u,x}P^{X}(x)\And P^{Y}(y),\\
P^{Y}(v)[P^{X}(x)\And P^{Y}(y)]&=&\de_{v,y}P^{X}(x)\And P^{Y}(y).
\eeqas
It follows that
\beqas
P^{X}(u)P^{Y}(v)\ket{\Psi}
&=&
\sum_{x,y}P^{X}(u)P^{Y}(v)[P^{X}(x)\And P^{Y}(y)]\ket{\Psi}\\
&=&
P^{X}(u)\And P^{Y}(v)\ket{\Psi}.
\eeqas
By symmetry we obtain
$$
P^{X}(u)P^{Y}(v)\ket{\Psi}=P^{Y}(v)P^{X}(u)\ket{\Psi}.
$$
Thus, (ii) holds.

(ii) $\Then$ (iii):
Let 
$$
\mu(x,y)=\av{\Ps|P^{X}(x)P^{Y}(y)|\Ps}.
$$
Then $\mu(x,y)\ge 0$ for all $x,y\in\R$,
since $$P^{X}(x)P^{Y}(y)\ket{\Psi}=P^{X}(y)P^{Y}(x)P^{X}(y)\ket{\Psi}$$
by assumption,
 and $\sum_{x,y}\mu(x,y)=1$.
Let $f(X,Y)=X^{n_1}Y^{m_1}\cdots X^{n_N}Y^{m_N}$ with 
$0\le n_1,m_1,\ldots,n_N,m_N$.
Then by assumption we have
\beqas
f(X,Y)\ket{\Psi}
&=&\sum_{x,y} x^{n_1+\cdots+ n_N}y^{m_1+\cdots+m_N} P^{X}(x)P^{Y}(y)\ket{\Psi},\\
f(x,y)&=&x^{n_1+\cdots+ n_N}y^{m_1+\cdots+m_N}.
\eeqas
Thus, we have
\beqas
\av{\Ps|f(X,Y)|\Ps}
&=&\sum_{x,y}x^{n_1+\cdots+ n_N}y^{m_1+\cdots+m_N}\mu(x,y)\\
&=&\sum_{x,y}f(x,y)\mu(x,y).
\eeqas
By linearity, the relation
$$
\av{\Ps|f(X,Y)|\Ps}=\sum_{x,y}f(x,y)\mu(x,y)
$$
holds for every polynomial $f(X,Y)$.

(iii) $\Then$ (i): Suppose that there exists a JPD $\mu(x,y)$ of $X$ and $Y$ in $\ket{\Psi}$.
Let $u,v\in \R$.  Let $f(X), g(Y)$ be polynomials of $X$ and $Y$.
We have
\beqas
\lefteqn{
\av{\Ps|[f(X),g(Y)]^{\da}[f(X),g(Y)]|\Ps}}\quad\\
&=&\sum_{x,y}|f(x)g(y)-g(y)f(x)|^2\mu(x,y)=0,
\eeqas
and hence
\[
[f(X),g(Y)]\ket{\Psi}=0.
\]
Taking $f(X), g(Y)$ as $f(X)=P^{X}(x)$ and $g(Y)=P^{Y}(y)$,
we have 
\[
P^{X}(x)P^{Y}(y)\ket{\Psi}=P^{Y}(y)P^{X}(x)\ket{\Psi},
\]
so that $X$ and $Y$ commute in $\ket{\Psi}$.
It follows that $P^{X}(x)P^{Y}(y)\ket{\Psi}$ is a common eigenvector 
of $X$ and $Y$ if $P^{X}(x)P^{Y}(y)\ket{\Psi}\ne0$.
It follows from $\ket{\Psi}=\sum_{x,y}P^{X}(x)P^{Y}(y)\ket{\Psi}$ that 
$\ket{\Psi}$ is a superposition of common eigenstate of $X$ and $Y$.

Suppose that (i)--(iv) hold and let
$\mu$ be a JPD of $X,Y$ in $\ket{\Psi}$.
Then 
$$
f(X,Y)P^{X}(x)P^{Y}(y)\ket{\Psi}=f(x,y)P^{X}(x)P^{Y}(y)\ket{\Psi}.
$$
It follows that 
$$
\av{\Ps|f(X,Y)|\Ps}=\sum_{x,y}f(x,y)\av{\Ps|P^{X}(x)P^{Y}(y)|\Ps}.
$$
From \Eq{JPD} we have
$$
\sum_{x,y}f(x,y)\av{\Ps|P^{X}(x)P^{Y}(y)|\Ps}
=
\sum_{x,y}f(x,y)\mu(x,y).
$$
Since $f(x,y)$ was arbitrary, we obtain
$$
\mu(x,y)=\av{\Ps|P^{X}(x)P^{Y}(y)|\Ps}.
$$
\end{Proof}

It should be noted that if 
\beql{JPD3}
\av{\Ps|P^{X}(x)P^{Y}(y)|\Ps}\ge 0
\eeq
for all $x,y\in\R$, then \Eq{JPD2} defines a probability distribution
$\mu(x,y)$ on $\R^2$ satisfying the marginal probability conditions:
\beqa
\mu(x,\R)&=&\av{\Psi|P^{X}(x)|\Psi},\\ 
\mu(\R,y)&=&\av{\Psi|P^{Y}(y)|\Psi},
\eeqa
where $\mu(x,\R)=\sum_y\mu(x,y)$ and $\mu(\R,y)=\sum_x\mu(x,y)$.
However, \Eq{JPD3} does not ensure that $\mu(x,y)$ 
satisfies \Eq{JPD}.
In fact,
let $X=\si_x$, $Y=\si_y$, and $\ket{\Psi}=\ket{\si_x=+1}$,
where $\si_x,\si_y$ are Pauli operators on $\C^2$.
Let $f(X,Y)=YXY$.  Then we have
\beqa
&\mu(+1,+1)=1/2, \quad \mu(+1,-1)=1/2,&\\
&\mu(-1,+1)=0,\quad \mu(-1,-1)=0,&\\
&\av{\Psi|f(X,Y)|\Psi}=\av{\Psi|-X|\Psi}=-1,&\\
&\sum_{x,y}f(x,y)\mu(x,y)=\sum_{x,y}xy^2\mu(x,y)=+1.&
\eeqa
Thus, \Eq{JPD} does not hold.
\medskip 

\paragraph*{\bf State-dependent definition for accurate measurements 
of quantum observables.}
To characterize accurate measurements of a quantum observable
in a given state, here, we take two approaches, one based on classical
correlation and the other based on quantum correlation,
which will be eventually shown to be equivalent.

As discussed before, if $A(0)$ and $M(\t)$ commute in $\ket{\psi,\xi}$,  
there exists the JPD $\mu(x,y)$ of $A(0)$ and $M(\t)$ in $\ket{\psi,\xi}$,
which describes the classical input-output correlation.
Then according to the consistency with the classical description,
the observable $A$ is considered to be accurately measured 
if  $A(0)$ and $M(\t)$ are perfectly correlated in their JPD $\mu$, 
i.e., $\mu(A(0)=M(\t))=1$.
Thus, we reach the following 
condition for the measuring process $\M$ to accurately measure $A$ in 
the state $\ket{\psi}$: 
\medskip

(S) $A(0)$ and $M(\t)$ commute in $\ket{\psi,\xi}$ and their JPD $\mu$ 
satisfies $\mu(A(0)=M(\t))=1$.
\medskip

In the second approach, we consider the {\em weak joint distribution (WJD)}
$\nu(x,y)$ of $A(0)$ and $M(\t)$ in $\ket{\psi,\xi}$ defined by
\beq
\nu(x,y)=\av{\psi,\xi|P^{M(\t)}(y)P^{A(0)}(x)|\psi,\xi}.
\eeq
From Theorem 1,
if $A(0)$ and $M(\t)$ commute in $\ket{\psi,\xi}$, the WJD $\nu(x,y)$
coincides with the JPD $\mu(x,y)$ of $A(0)$ and $M(\t)$ in $\ket{\psi,\xi}$.
The WJD always exists, and is operationally accessible 
by weak measurement and post-selection
\cite{LW10},  but possibly takes negative or complex values.
Then it is natural to consider the following condition:
\medskip

(W) The WJD of $A(0)$ and $M(\t)$ in $\ket{\psi,\xi}$ satisfies $\nu(x,y)=0$ 
if $x\ne y$.
\medskip

Since the WJD is operationally accessible, condition (W) is also operationally accessible.
Obviously, (W) is logically weaker than or equivalent to (S). 
If condition (S) holds, the measurement should be considered 
an accurate measurement 
for the consistency with the classical description.
On the other hand, if the measurement is accurate, 
any operational test for the possible error should be passed.
Observing the WJD is one of available tests for 
the accurate measurement, and it is natural to consider that the test is passed
if  $\nu(x,y)=0$ for all $x,y$ with $x\ne y$ and that the test is failed,
or the error is witnessed, if $\nu(x,y)\not=0$ for some $x,y$ with $x\ne y$;
this type of test has been discussed in detail by Mir et al. \cite{MLMSGW07} 
and Garretson et al. \cite{GWPP04} in the context of witnessing momentum 
transfer in a which-way measurement.
Thus, condition (W) should be satisfied by any accurate measurement, 
since a failure of (W), or a non-zero value of $\nu(x,y)$ for a pair $(x,y)$, 
witnesses an error of the measurement.

Therefore, condition (S) is a sufficient condition for the measurement
to be accurate, and condition (W) a necessary condition.
The following theorem shows that both conditions are actually equivalent 
so that both of them are necessary and sufficient conditions 
for the measurement to be accurate.

\begin{Theorem}
For any measuring process $\M$, an observable $A$, and a state $\ket{\psi}$,
condition (S) and condition (W) are equivalent.
\end{Theorem}
\begin{Proof}
The assertion was generally proved in Ref.~\cite{05PCN,06QPC}
after a lengthy argument.  Here, we give a direct proof.
Since (S) implies (W), it suffices to show the implication (W)$\Then$(S).
Suppose that the WJD $\nu(x,y)$ of $A(0)$ and $M(\t)$ in $\ket{\psi,\xi}$
satisfies $\nu(x,y)=0$ if $x\ne y$.
Then
\beqas
\av{\psi,\xi|P^{A(0)}(x)P^{M(\t)}(x)|\psi,\xi}
&=&\av{\psi,\xi|P^{A(0)}(x)|\psi,\xi},\\
\av{\psi,\xi|P^{A(0)}(x)P^{M(\t)}(x)|\psi,\xi}
&=&\av{\psi,\xi|P^{M(\t)}(x)|\psi,\xi}.
\eeqas
Consequently,
\beqas
\|P^{A(0)}(x)\ket{\psi,\xi}-P^{M(\t)}(x)\ket{\psi,\xi}\|^2
=0,
\eeqas
and 
\beqas
P^{A(0)}(x)\ket{\psi,\xi}=P^{M(\t)}(x)\ket{\psi,\xi}
\eeqas
Thus, 
\beqas
P^{A(0)}(x)P^{M(\t)}(y)\ket{\psi,\xi}&=&\de_{x,y}P^{A(0)}(x)\ket{\psi,\xi},\\
P^{M(\t)}(y)P^{A(0)}(x)\ket{\psi,\xi}&=&\de_{x,y}P^{A(0)}(x)\ket{\psi,\xi}.
\eeqas
It follows that $A(0)$ and $M(\t)$ commute in $\ket{\psi,\xi}$ and the condition in
$(S)$ holds,
Thus the implication (W)$\Then$(S) follows.
\end{Proof}
\medskip 

\paragraph*{\bf  Sound and complete quantum root-mean-square errors.}
In addition to the locally uniform q-rms error, here, we introduce the following
two sorts of q-rms errors.
For any invertible density function $f$, we define the
{\em $f$-distributed q-rms error} $\ep_{f}$ by
\beql{f-q-rms}
\ep_{f}(A,\Pi,\ket{\psi})^2=\int_{\R}\ep_{t}(A,\Pi,\ket{\psi})^2 f(t) \,dt.
\eeq
For any invariant mean $m$ on $\R$ \cite{Sri80}, define the {\em $m$-distributed q-rms error} 
$\ep_{m}$ by
\beql{m-q-rms}
\ep_{m}(A,\Pi,\ket{\psi})^2=m_t[\ep_{t}(A,\Pi,\ket{\psi})^2].
\eeq

Then we have the following theorem.
\begin{Theorem}
 The following statements hold.

(i) Error measures $\epu$, $\ep_f$, and $\ep_m$ are sound and complete q-rms errors.

(ii) Error measure $\epu$  has  the dominating property, (V).

(iii) Error measure $\epu$, $\ep_f$, and $\ep_m$ have the conservation property for dichotomic measurements, (VI).

(iv) The relations 
\[
\ep_m\le\epu,\quad  \sup_{f}\ep_f=\epu,
\]
hold for any invariant mean $m$, where $f$ varies over 
all the invertible density functions.

(v) Error measure $\ep_m$ satisfies the relation
\begin{align*}
\ep_{m}(A,\Pi,\ket{\psi})^2=\ep_{NO}\left(A,\Pi,\sum_n P^{A}(a_n)\ketbra{\psi}P^{A}(a_n)\right)^2
\end{align*} 
if  $A=\sum_{n}a_n P^{A}(a_n)$.
\end{Theorem}
\begin{Proof}
It is obvious from definition that $\epu$ satisfies the operational definability, (I).
From \Eq{ept-epn}, $\epu$ satisfies the correspondence principle,  (II),
and hence satisfies the soundness, (III).
To prove the completeness, (IV),
suppose $\overline{\ep}(A,\Pi,\ket{\psi})=0$.
Then we have
$$
M(\t)e^{-itA}{\ket{\psi}}\ket{\xi}=A(0)e^{-itA}{\ket{\psi}}{\ket{\xi}}.
$$
Since $t$ was arbitrary, we have
\beqas
M(\t)\sum_{j}a_j e^{-it_jA}{\ket{\psi}}\ket{\xi}=A(0)\sum_{j}a_j e^{-it_jA}{\ket{\psi}}\ket{\xi}
\eeqas
for any $\{a_j\}$ and $\{t_j\}$.
By Fourier expansion,  
the set of operators $\sum_{j}a_j e^{-it_jA}$ includes all functions of $A$,
so that we have 
\beqas
M(\t)P^{A}(x){\ket{\psi}}\ket{\xi}
&=&A(0)P^{A}(x){\ket{\psi}}\ket{\xi}\\
&=&xP^{A}(x){\ket{\psi}}\ket{\xi}
\eeqas
for all $x\in \R$.
Thus, $P^{A}(x){\ket{\psi}}\ket{\xi}$ is a common eigenstate of $M(\t)$
and $A(0)$ if  $P^{A}(x){\ket{\psi}}\ne 0$, 
and ${\ket{\psi}}\ket{\xi}=\sum_{x}P^{A}(x){\ket{\psi}}\ket{\xi}$
 is a superposition of common eigenstates of $M(\t)$
and $A(0)$.
Thus, $\epu$ satisfies the completeness requirement (IV). 
Therefore, we conclude that $\epu$ is a complete q-rms error.
The proofs for  $\ep_f$ and $\ep_m$ are similar.

Assertion (ii) follows immediately from the definition.

To prove assertion (iii), suppose  $A^2=\hat{\Pi}^{(2)}=I$.
Let $\ket{\psi_t}=e^{-itA}\ket{\psi}$.
Then we have the commutation relation
\beqas
[A\,\hat{\Pi}+\hat{\Pi}A,A]=0,
\eeqas
and hence 
\beqas
\lefteqn{2\Re\av{\psi_t|A\,\hat{\Pi}|{\psi_t}}}\\
&=&\av{\psi_t|(A\hat{\Pi}+\hat{\Pi}A)|{\psi_t}}\\
&=&\av{\psi|(A\hat{\Pi}+\hat{\Pi}A)|\psi}\\
&=&2\Re\av{\psi|A\,\hat{\Pi}|\psi}.
\eeqas
Thus, we have
\beqas
\lefteqn{
\ep_t(A,\Pi,\ket{\psi})}\\
&=&\av{\psi_t|A^2|\psi_t}+\av{\psi_t|\hat{\Pi}^{(2)}|\psi_t}-2\Re\av{\psi_t|A\,\hat{\Pi}|\psi_t}\\
&=&\av{\psi|A^2|\psi}+\av{\psi|\hat{\Pi}^{(2)}|\psi}-2\Re\av{\psi|A\,\hat{\Pi}|\psi}\\
&=&\epn(A,\Pi,\ket{\psi}).
\eeqas
Thus, assertion (iii) follows. 

Assertion (iv) follows easily from the properties of integral and
invariant mean.

Assertion (v) follows from \cite[Theorem 5.2]{Sri80}.
\end{Proof}
Consider a quantum system with single degree of freedom described 
by a pair of canonically conjugate observables $Q,P$ prepared in
the state $\ket{\ph}$ such that $|\bracket{p|\ph}|^2=f(p)$.
By the relation
\begin{align*}
\lefteqn{
\int_{\R}\ep_{t}(A,\Pi,\ket{\psi})^2 f(t) dt}\\
&=
\ep_{NO}\left(A,\Pi,\int_{\R}e^{-ipA}\ketbra{\psi}e^{ipA}|\bracket{p|\ph}|^2\,dt\right)^2,
\end{align*}
the above definition of $\ep_{f}$ is equivalent to making the canonical 
approximate
$A$-measurement with the $Q$-meter prepared in 
the state $\ket{\ph}$ such that $|\bracket{p|\ph}|^2=f(p)$ in the $P$-basis
before evaluating the noise-operator-based rms error \cite{88MR,93CA} .
The definition of $\ep_{m}$ is also equivalent to making the canonical 
approximate
$A$-measurement with the $Q$-meter prepared in the $m$-Dirac
state before evaluating the noise-operator-based rms error \cite{88MR}.
It is well-known that there is no canonical choice of $f$ or $m$ in general 
to achieve the ideal measurement of an arbitrary $A$ \cite{Sri80,88MR}. 
By Theorem 3 (iv), our definition for $\epu$ is equivalent to
\[
\epu(A,\Pi,\ket{\psi})=\sup_f \ep_{f}(A,\Pi,\ket{\psi}),
\]
where $f$ varies over all the invertible wave functions.
Thus, although there is no canonical choice of $f$ in general, 
the definition of $\epu$ can be interpreted as choosing 
the most error-sensitive $f$ among all the invertible wave functions $f$.
\bigskip

\paragraph*{\bf Data Availability.}
Data sharing not applicable to this article as no datasets were generated 
or analysed during the current study.

\paragraph*{\bf Acknowledgments.}
The author acknowledges the support of the JSPS KAKENHI, No.~26247016 
and No.~17K19970, and of the IRI-NU collaboration.

\paragraph*{\bf Competing Interests.}
The author has no competing interests relevant to this work.

\paragraph*{\bf Author Contributions.}
The author researched, collated, and wrote this paper.

\end{document}